\documentclass[twocolumn,showpacs,preprintnumbers,amsmath,amssymb,prb]{revtex4}

\usepackage{graphicx}
\usepackage{dcolumn}
\usepackage{bm}
\usepackage{amssymb}

\begin{document}


\title{Metal-insulator transition in graphene induced by circularly polarized photons}

\author{O.V. Kibis}\email{Oleg.Kibis@nstu.ru}

\affiliation{Department of Applied and Theoretical Physics,
Novosibirsk State Technical University, Karl Marx Avenue 20,
630092 Novosibirsk, Russia}


\begin{abstract}
Exact stationary solutions of the electron-photon Dirac equation
are obtained to describe the strong interaction between massless
Dirac fermions in graphene and circularly polarized photons. It
follows from them that this interaction forms bound
electron-photon states which should be considered as a kind of
charged quasiparticles. The energy spectrum of the quasiparticles
is of dielectric type and characterized by an energy gap between
the valence and conductivity bands. Therefore the electron-photon
interaction results in metal-insulator transition in graphene. The
stationary energy gap, induced by photons, and concomitant effects
can be observed for graphene exposed to a laser-generated
circularly polarized electromagnetic wave.
\end{abstract}


\maketitle
\section{INTRODUCTION}
Since the discovery of graphene \cite{Novoselov_05} --- a
monolayer of carbon atoms --- its unique physical characteristics
have aroused enormous interest in the scientific community.
Particularly, the influence of electromagnetic field on electronic
properties of graphene is in focus of attention
\cite{CastroNeto_09}. However all theoretical models, elaborated
to describe this influence, deal with strong electromagnetic field
in the framework of classical electrodynamics. A theory, capable
of describing the strong interaction between electrons in graphene
and quantized electromagnetic field, was unknown up to the
present. The paper is aimed to fill partially this gap in the
theory. It is surprising that the problem of interaction between
electrons at the Fermi level of graphene and circularly polarized
photons can be solved exactly in analytical form without the need
of any approximation and numerical calculations. This is one of
few exactly solvable problems in the theory of photon-matter
interaction, that is not without interest to both condensed matter
physics and quantum optics. Moreover, since electrons in graphene
behave as massless relativistic fermions
\cite{Wallace_47,Lukyanchuk_04,Novoselov_05}, graphene can serve
as a low-energy `proving ground' for the relativistic physics
\cite{Semenoff_84}. Therefore the obtained solutions can also be
useful to describe the interaction between relativistic particles
and photons. Thus the solved problem lies at the intersection of
different excited fields of modern physics, that is what caused to
write this paper.
\section{DIRAC PROBLEM FOR ELECTRON-PHOTON SYSTEM IN GRAPHENE}
Generally, electron states in graphene near the Fermi energy are
described by eight-component wave functions written in a basis
corresponding to two crystal sublattices of graphene, two electron
valleys, and two orientations of electron spin
\cite{CastroNeto_09}. In what follows intervalley scattering
processes and spin effects will be beyond consideration, that
reduces the number of necessary wave-function components to two.
To describe the interaction between a plane monochromatic
circularly polarized electromagnetic wave and electrons in a
graphene, we shall be to use the Cartesian coordinate system
$(x,y,z)$, where the axis $z$ is perpendicular to the graphene
fixed at $z=0$. Then near the point at the Fermi level, where the
valence and conductivity bands of graphene touch each other (the
Dirac point), the Hamiltonian of electrons interacting with the
electromagnetic field can be written for a single valley and for a
certain direction of electron spin in the form
\cite{CastroNeto_09}
\begin{equation}\label{E}
\hat{{\cal H}}_e=
v_F\hat{\bm{\sigma}}\left(\hat{\mathbf{p}}-\frac{e}{c}\mathbf{A}\right)\,,
\end{equation}
where $v_F$ is Fermi velocity, $\hat{\mathbf{p}}$ is operator of
electron momentum in the graphene plane, $e$ is electron charge,
and $\mathbf{A}$ is vector potential of the electromagnetic field.
As to the vector operator $\hat{\bm{\sigma}}$, its components are
Pauli matrices written in the basis of two orthogonal electron
states arisen from two crystal sublattices of graphene. Since the
operator $\hat{{\sigma}}_z$ is diagonal, these two states can be
denoted by mutually opposite orientations of the pseudospin along
the $z$-axis, $s_z=\pm1/2$. Considering the problem within the
standard quantum-field approach \cite{Landau_4}, the classical
field, $\mathbf{A}$, should be replaced with the field operator,
$\hat{\mathbf{A}}$. Assuming the electromagnetic wave to be
clockwise-polarized and propagating along the $z$-axis, this
operator can be written as
\begin{equation}\label{A}
\hat{\mathbf{A}}=\sqrt{2\pi\hbar
c^2/\omega_0V}\left(\mathbf{e}_+\hat{a}+
\mathbf{e}_-\hat{a}^\dagger\right)\,,
\end{equation}
where $\omega_0$ is frequency of the electromagnetic wave, $V$ is
volume of space with the field,
$\mathbf{e}_{\pm}=(\mathbf{e}_{x}\pm i\mathbf{e}_{y})/\sqrt{2}$
are polarization vectors, $\mathbf{e}_{x,y}$ are unit vectors
directed along the $x,y$-axes, $\hat{a}$ and $\hat{a}^\dagger$ are
photon operators of annihilation and creation, respectively,
written in the Schr\"{o}dinger representation (the representation
of occupation numbers) \cite{Landau_4}. Then the complete
Hamiltonian of the electron-photon system, including both the
field energy, $\hbar\omega_0\hat{a}^\dagger\hat{a}$, and the
electron Hamiltonian (\ref{E}) rewritten in the quantum-field
form, is given by
\begin{equation}\label{H}
\hat{{\cal H}}=\hbar\omega_0\hat{a}^\dagger\hat{a}+
v_F\hat{\bm{\sigma}}\hat{\mathbf{p}}-e\sqrt{\frac{4\pi\hbar v_F^2}
{\omega_0V}}\left(\hat{\sigma}_+\hat{a}+\hat{\sigma}_-\hat{a}^\dagger
\right)\,,
\end{equation}
where $\hat{\sigma}_{\pm}=(\hat{\sigma}_x\pm i\hat{\sigma}_y)/2$
are step-up and step-down operators for the $z$-projection of the
pseudospin.

The stationary solutions of the effective Dirac equation with the
Hamiltonian (\ref{H}) have the form
$\Psi(\mathbf{k})=e^{i\mathbf{k}\mathbf{r}}\psi(\mathbf{k})$,
where $\mathbf{k}$ is wave vector in the graphene plane,
$\mathbf{r}$ is radius-vector, and $\psi(\mathbf{k})$ is
eigenstate of the Hamiltonian
\begin{equation}\label{HK}
\hat{{\cal H}}_{\mathbf{k}}=\hbar\omega_0\hat{a}^\dagger\hat{a}+
\hbar v_F\hat{\bm{\sigma}}{\mathbf{k}}-e\sqrt{\frac{4\pi\hbar
v_F^2}
{\omega_0V}}\left(\hat{\sigma}_+\hat{a}+\hat{\sigma}_-\hat{a}^\dagger
\right)\,.
\end{equation}
At the Dirac point ($\mathbf{k}=0$) the Hamiltonian (\ref{HK}) is
formally similar to the Hamiltonian of exactly solvable
Jaynes-Cummings model \cite{Jaynes_63}. As a consequence, the
effective Dirac equation with the Hamiltonian (\ref{HK}) for
$\mathbf{k}=0$ can also be solved exactly. To describe the
electron-photon system at the Dirac point, let us use the notation
$|s_z,N\rangle$ which indicates that the electron is in one of two
quantum states with the pseudospin projections $s_z=\pm1/2$ and
the electromagnetic field is in quantum state with the photon
occupation number $N=1,2,3,...$. Then exact eigenstates of the
Hamiltonian (\ref{HK}) for $\mathbf{k}=0$,
$\psi(0)=\varphi_{+\frac{1}{2},N}$ and
$\psi(0)=\varphi_{-\frac{1}{2},N}$, can be written as
\begin{eqnarray}\label{phi}
|\varphi_{\pm\frac{1}{2},N}\rangle&=&
\sqrt{\frac{\Omega_\pm+\omega_0}{2\,\Omega_\pm}}\,|\pm1/2,N\rangle\nonumber\\
&\pm&\frac{e}{|e|}\sqrt{\frac{\Omega_\pm-\omega_0}{2\,\Omega_\pm}}\,|\mp1/2,N\pm1\rangle\,,
\end{eqnarray}
where $$\Omega_\pm=\sqrt{16(N+1/2\pm1/2)(\pi
v_F^2e^2/\hbar\omega_0V)+\omega_0^2}\,,$$ energies of the
electron-photon states (\ref{phi}), $\varepsilon_{+\frac{1}{2},N}$
and $\varepsilon_{-\frac{1}{2},N}$, are given by
\begin{equation}\label{Energy}
\varepsilon_{\pm\frac{1}{2},N}=
N\hbar\omega_0\pm\frac{\hbar\omega_0}{2}\mp\frac{\hbar\Omega_\pm}{2}\,,
\end{equation}
and subscript indices in Eqs.~(\ref{phi})--(\ref{Energy}) indicate
genesis of these states: The state
$|\varphi_{\pm\frac{1}{2},N}\rangle$ turns into the state
$|\pm1/2,N\rangle$ when the electron-photon interaction vanishes
(i.e. for $e=0$). The expressions (\ref{phi})--(\ref{Energy}) can
be easily verified by direct substitution into the effective Dirac
equation $\hat{{\cal
H}}_{\mathbf{k}}\varphi_{\pm\frac{1}{2},N}=\varepsilon_{\pm\frac{1}{2},N}\varphi_{\pm\frac{1}{2},N}$
with the Hamiltonian (\ref{HK}) for $\mathbf{k}=0$, keeping in
mind the trivial relations \cite{Landau_3,Landau_4}
\begin{eqnarray}
\hat{\sigma}_{\pm}\,|\mp1/2,N\rangle&=&|\pm1/2,N\rangle\,,\,\,\,\,
\hat{\sigma}_{\pm}\,|\pm1/2,N\rangle\,=0\,\,,\nonumber\\
\hat{a}^{\dagger}\,|\pm1/2,N\rangle&=&\sqrt{N+1}\,|\pm1/2,N+1\rangle\,,\nonumber\\
\hat{a}\,|\pm1/2,N\rangle&=&\sqrt{N}\,|\pm1/2,N-1\rangle\,.\nonumber
\end{eqnarray}
Let the photon occupation number of the electromagnetic field
(\ref{A}) is fixed at $N=N_0$ by a field source. Then the
electron-photon states $\varphi_{-\frac{1}{2},N_0}$ and
$\varphi_{+\frac{1}{2},N_0}$, originated from electron states
degenerated at the Dirac point, are separated by the energy gap
$\varepsilon_g=\varepsilon_{-\frac{1}{2},N_0}-\varepsilon_{+\frac{1}{2},N_0}$.
In what follows we shall be to assume the electromagnetic wave to
be classically strong, that corresponds to macroscopically large
photon occupation numbers ($N_0\gg1$). Then the gap can be written
as
\begin{equation}\label{Gap}
\varepsilon_g=\sqrt{W_0^2+(\hbar\omega_0)^2}-\hbar\omega_0\,,
\end{equation}
where $W_0=2v_FeE_0/\omega_0$ is energy of electron rotational
motion induced by the circularly polarized wave, and
$E_0=\sqrt{4\pi N_0\hbar\omega_0/V}$ is classical amplitude of
electric field of the wave.

Let us proceed to solving the effective Dirac equation with the
Hamiltonian (\ref{HK}) near the Dirac point for wave vectors
$\mathbf{k}\neq0$. Since the obtained electron-photon states
(\ref{phi}) are eigenstates of the same Hamiltonian (\ref{HK}) for
$\mathbf{k}=0$, they form complete basis of the considered
electron-photon system. Therefore true eigenstates of the
Hamiltonian (\ref{HK}) for $\mathbf{k}\neq0$ can be sought as an
expansion
\begin{equation}\label{psi}
\psi(\mathbf{k})=\sum_{m=1}^{\infty}\left[c_{+\frac{1}{2},m}(\mathbf{k})\varphi_{+\frac{1}{2},m}+
c_{-\frac{1}{2},m}(\mathbf{k})\varphi_{-\frac{1}{2},m}\right].
\end{equation}
Substituting the expansion (\ref{psi}) into the effective Dirac
equation $\hat{{\cal
H}}_{\mathbf{k}}\psi(\mathbf{k})=\varepsilon(\mathbf{k})\psi(\mathbf{k})$
with the Hamiltonian (\ref{HK}), we arrive at the system of
recurrent algebraic equations for coefficients
$c_{\pm\frac{1}{2},m}(\mathbf{k})$. This system can be easily
solved in the special case, when the effective parameter of
electron-field interaction, $\alpha=W_0/\hbar\omega_0$, satisfies
the condition $\alpha\ll1$. Seeking the dispersion dependencies
$\varepsilon(\mathbf{k})$ and $c_{\pm\frac{1}{2},m}(\mathbf{k})$
for any $\mathbf{k}$ in the principal order with respect to
$\alpha$, the system of recurrent equations can be reduced to the
two equations
\begin{eqnarray}
\left(\varepsilon-N_0\hbar\omega_0-\frac{\varepsilon_g}{2}\right)c_{-\frac{1}{2},N_0}
&=&\hbar v_F(k_x+ik_y)c_{+\frac{1}{2},N_0}\,,\,\,\,\,\,\,\,\,\,\label{AE1}\\
\left(\varepsilon-N_0\hbar\omega_0+\frac{\varepsilon_g}{2}\right)c_{+\frac{1}{2},N_0}
&=&\hbar
v_F(k_x-ik_y)c_{-\frac{1}{2},N_0}\,,\,\,\,\,\,\,\,\,\,\label{AE2}
\end{eqnarray}
which can be solved exactly. Solving Eqs.~(\ref{AE1})--(\ref{AE2})
by conventional methods, we obtain two branches of the energy
spectrum, $\varepsilon(\mathbf{k})$, which can be written as
\begin{equation}\label{EnergyK}
\varepsilon^\pm_{N_0}(\mathbf{k})=N_0\hbar\omega_0\pm\sqrt{(\varepsilon_g/2)^2+(\hbar
v_Fk)^2}\,,
\end{equation}
where omitted $\mathbf{k}$-terms are $\sim o(\alpha)$. As a
result, eigenstates of the Hamiltonian (\ref{H}) are given by
\begin{equation}\label{Psi}
\Psi^\pm_{N_0}(\mathbf{k})=e^{i\mathbf{k}\mathbf{r}}\left[c^\pm_{+\frac{1}{2},N_0}(\mathbf{k})
\varphi_{+\frac{1}{2},N_0}
+c^\pm_{-\frac{1}{2},N_0}(\mathbf{k})\varphi_{-\frac{1}{2},N_0}\right]\,,
\end{equation}
where superscript indexes `$\pm$' denote solutions of
Eqs.~(\ref{AE1})--(\ref{AE2}) corresponding to the two energy
branches (\ref{EnergyK}). As expected, for $\mathbf{k}=0$ the
obtained expressions (\ref{EnergyK})--(\ref{Psi}) coincide with
the exact solutions (\ref{Energy})--(\ref{phi}).
\section{ELECTRON-PHOTON QUASIPARTICLES IN GRAPHENE AND THEIR PROPERTIES}
The full energy of the electron-photon system (\ref{EnergyK})
consists of the field energy, $N_0\hbar\omega_0$, and the term
\begin{equation}\label{EK}
\Delta\varepsilon(\mathbf{k})=\pm\sqrt{(\varepsilon_g/2)^2+(\hbar
v_Fk)^2}\,,
\end{equation}
which is arisen from electron-photon interaction. As a
consequence, an electron interacting with circularly polarized
photons can be considered formally as a quasiparticle with the
energy spectrum (\ref{EK}). From the viewpoint of classical
electrodynamics, the circularly polarized electromagnetic wave
rotates the electron \cite{Landau_2}. Therefore the quasiparticle
is the mix of photon states and electron states corresponding to
this rotational motion of the electron. Certainly, a center of
trajectory of the rotating electron can move in the graphene
plane, that corresponds to a translational motion of the
quasiparticle as a whole. The translational motion is described in
Eq.~(\ref{EK}) by the wave vector of quasiparticle, $\mathbf{k}$.
Since the quasiparticle is {\it rotating} electron dressed by
circularly polarized photons, this wave vector differs from the
wave vector of free electron in graphene.

The energy spectrum of quasiparticles (\ref{EK}), schematically
pictured in Fig.~1a, defines optical and transport properties of
graphene irradiated by circularly polarized light.
\begin{figure}[th]
\includegraphics[width=0.48 \textwidth]{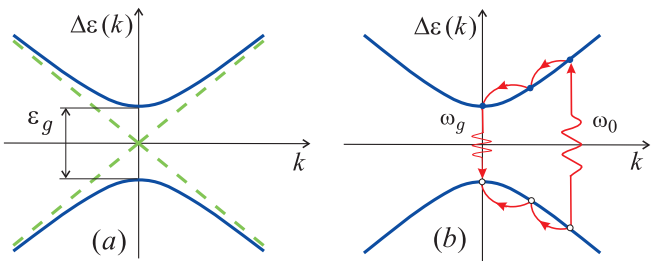}
\caption{(Color online) (a) Energy spectrum of quasiparticles in
graphene in the presence of circularly polarized irradiation
(solid line) and in the absence of one (dashed line); (b) Scheme
of quasiparticle transitions in graphene irradiated at the
frequency $\omega_0$.}\label{fig1}
\end{figure}
The distinctive feature of the spectrum (\ref{EK}) is the energy
gap (\ref{Gap}) between quasiparticle states originated from
valence and conductivity bands of graphene. Formally, the value of
the photon-induced gap in the limit of large radiation
intensities, given by Eq.~(\ref{Gap}), is the same as the value of
dynamical gap induced at the Dirac point of graphene by a
classical circularly polarized electromagnetic wave
\cite{Oka_09,Oka_09_1}. However, there is a conceptual difference
between the {\it stationary} energy gap (\ref{Gap}) arisen from
quantum-field theory and {\it dynamical} gaps
\cite{Oka_09,Oka_09_1,Lopez_08,Efetov_08} arisen in graphene from
classical fields. Historically, the term `dynamical gap' is come
into the theory of graphene-field interaction
\cite{Oka_09,Oka_09_1,Lopez_08,Efetov_08} from the general theory
\cite{DittrichBook,Kohler_05} developed to describe different
quantum systems exposed to time-dependent classical fields and
denotes the gap in spectrum of Floquet quasienergies. Thus the
dynamical gap is not true gap in the density of states of charge
carriers. As it was specially stressed in
Ref.~\onlinecite{Lopez_08}, ``since it is time-dependent problem,
one cannot measure the gap directly from the density of states''.
In contrast to the dynamical gap, the gap (\ref{Gap}) is obtained
from stationary solutions of the time-independent Dirac problem
for the electron-photon system and is true gap in the density of
bound electron-photon states (charged quasiparticles). As a
consequence, the stationary gap (\ref{Gap}) will be to manifest
itself directly in all phenomena sensitive to the density of
states of charge carriers. In other words, the incorporation of
such an additional physical factor as a quantum nature of
electromagnetic field into the theory of graphene-field
interaction, presented in the given paper, allows to consider the
strong interacting electron-photon system in graphene as a gas of
noninteracting quasiparticles with density of states defined by
the energy spectrum (\ref{EK}). Since the quasiparticle concept is
conventional approach of the modern physics to describe quantum
systems with a strong interaction, the developed theory is
fruitful to predict and analyze different unexplored phenomena in
graphene irradiated by light --- particularly, novel physical
effects discussed hereafter.

If another electromagnetic field with the frequency $\omega$ is
applied to the considered electron-photon system, optical
transitions of quasiparticles between its valence and conductivity
bands can be possible. It follows from the energy conservation law
that the transitions are allowed for $\omega\geq\omega_g$, where
$\omega_g=\varepsilon_g/\hbar$. Therefore the gap (\ref{Gap})
results in the threshold of absorption of an external field at the
frequency $\omega=\omega_g$. Since the threshold frequency,
$\omega_g$, is controlled by the field intensity, $E_0$, this
effect can be used as a basis for tunable detectors of
electromagnetic radiation. The transition rate at the threshold
frequency is defined by the interband dipole matrix element
\begin{equation}\label{d}
|\langle\Psi_{N_0}^+(0)|\hat{d}_i|\Psi_{N_0}^-(0)\rangle|=
|e|\hbar
v_F\frac{\varepsilon_g+2\hbar\omega_0}{2\varepsilon_g(\varepsilon_g+\hbar\omega_0)}\,,
\end{equation}
where $\hat{\mathbf{d}}$ is dipole moment operator, and $i=x,y$.
Besides optical transitions induced by the external field, there
is the quasiparticle transition at the frequency $\omega_0$,
\begin{equation}\label{t}
\Psi_{N_0}^-(\mathbf{k})\rightarrow\Psi_{N_0-1}^+(\mathbf{k})\,,
\end{equation}
which is accompanied by one-photon absorption of the field
(\ref{A}). This transition, pictured in Fig.~1b, needs a special
discussion since the field (\ref{A}) is inherent in the solved
Dirac problem and cannot be considered as an external field. It
follows from Eq.~(\ref{EnergyK}) that for $\omega_g\leq\omega_0$
there is the wave vector, $\mathbf{k}^\prime$, satisfying the
condition
$\varepsilon_{N_0}^-(\mathbf{k}^\prime)=\varepsilon_{N_0-1}^+(\mathbf{k}^\prime)$.
Therefore the transition (\ref{t}) is allowed by the energy
conservation law for $\omega_g\leq\omega_0$ and can take place at
$\mathbf{k}=\mathbf{k}^\prime$ without the assistance of external
energy sources. However the field (\ref{A}) has been accounted
exactly by the complete Hamiltonian of the electron-photon system
(\ref{H}) and cannot lead directly to transitions between
different stationary states of the same Hamiltonian. To realize
the transition (\ref{t}), the eigenstates of the Hamiltonian
(\ref{H}), $\Psi_{N_0}^-(\mathbf{k})$ and
$\Psi_{N_0-1}^+(\mathbf{k})$, should be mixed by an external
elastic scatterer (for example, by a defect of graphene lattice).
Using Eq.~(\ref{Psi}), we can calculate the matrix element
\begin{equation}\label{matr}
\langle\Psi_{N_0-1}^+(\mathbf{k})\left|U(\mathbf{r})\right|\Psi_{N_0}^-(\mathbf{k})\rangle\,,
\end{equation}
which defines the rate of transition (\ref{t}) for any scatterring
field $U(\mathbf{r})$. In the absence of scatterers, when
$U(\mathbf{r})=0$ (or, generally, $U(\mathbf{r})=\mbox{const}$),
the matrix element (\ref{matr}) is zero and the transition is
impossible. Thus the quasiparticle transition (\ref{t}) is
scatterer-assisted process.

It is seen in Fig.~1b that excited quasiparticles, accomplishing a
cascade of non-radiative thermalization transitions after the
transition (\ref{t}), can create a population inversion between
the conductivity band and the valence band at $\mathbf{k}=0$.
Since the interband dipole moment (\ref{d}) is nonzero, that
results in light emission from graphene at the threshold
frequency, $\omega_g$, controlled by the field intensity, $E_0$.
Therefore the scheme, pictured in Fig.~1b, forms a physical basis
for creating tunable sources of electromagnetic radiation. The
proposed mechanisms for tunable detection and generation of
electromagnetic radiation are especially challenging for the THz
frequency range. Indeed, search for effective THz sources and
detectors is one of most excited problems of modern applied
physics \cite{Ferguson_02,Lee_07}. Moreover, the latest trend is
the using of different nanostructures to fill the THz gap
\cite{Dragoman_04,Kibis_07_NL,Portnoi_08,Batrakov_09,Kibis_09}.
Therefore the application of graphene for THz devices fits well
current tendencies in the graphene-based quantum electronics
\cite{Dragoman_09} and the nanophotonics as a whole.

Formally, the energy spectrum (\ref{EK}) is of dielectric type for
any field intensity, $E_0$. However the interband transition
(\ref{t}) leads to generation of free charge carriers in valence
and conductivity bands for $\omega_g\leq\omega_0$. To turn
graphene into true insulator, we need to forbid the transition
(\ref{t}) by the energy conservation law. This forbidding
corresponds to $\omega_g>\omega_0$. Therefore the metal-insulator
transition in graphene is a threshold effect and occurs at the
critical field intensity, $E_0=E_{c}$, satisfying the condition
$\varepsilon_g=\hbar\omega_0$. Using Eq.~(\ref{Gap}), we obtain
\begin{equation}\label{MIT}
E_c=\frac{\sqrt{3}\hbar\omega_0^2}{2v_F|e|}\,.
\end{equation}
Being in the insulator state for $E_0>E_{c}$, graphene has zero
conductivity for the temperature $T=0$ and behaves as a
semiconductor for $T\neq0$. Applying the conventional theory of
semiconductors \cite{AnselmBook} to graphene for $E_0>E_{c}$ and
$T\ll\varepsilon_g$, we can write the total density of free charge
carriers for a single valley of graphene and for a certain
direction of electron spin in the form
\begin{equation}\label{c}
n=\left(\frac{m^\ast
T}{\pi\hbar^2}\right)\exp\left(-\frac{\varepsilon_g}{2T}\right)\,,
\end{equation}
where $m^\ast$ is effective mass of quasiparticles, which defines
their energy spectrum,
\begin{equation}\label{EK2}
\Delta\varepsilon(\mathbf{k})=\pm\varepsilon_g/2\pm\hbar^2\mathbf{k}^2/2m^\ast\,,
\end{equation}
for small quasiparticle wave vectors $\mathbf{k}$. Considering
$\mathbf{k}$-terms in the Hamiltonian (\ref{HK}) as a perturbation
and using the standard perturbation theory \cite{Landau_3}, we can
find the energy spectrum of quasiparticles near their band edge as
a $\mathbf{k}$-power series expansion without assuming weakness of
the field (the parameter of electron-field interaction,
$\alpha=W_0/\hbar\omega_0$, is allowed to be not small). Obtaining
terms $\sim\mathbf{k}^2$ in the spectrum, we arrive at the exact
expression for the effective mass
\begin{equation}\label{m}
m^\ast=\frac{2\varepsilon_g(\varepsilon_g+2\hbar\omega_0)(\varepsilon_g+\hbar\omega_0)^2}
{v_F^2[(\varepsilon_g+2\hbar\omega_0)^3+\varepsilon_g^3]}\,.
\end{equation}
It should be noted that the energy spectrums of quasiparticles
(\ref{EK}) and (\ref{EK2}) are mutually complementary:
Eq.~(\ref{EK}) is applicable within a broad range of wave vectors
but only for small field intensities ($\alpha\ll1$), and
Eq.~(\ref{EK2}) with the effective mass (\ref{m}) is valid for any
field intensities but only for small wave vectors ($\hbar
v_Fk/\varepsilon_g\ll1$). As expected, Eqs. (\ref{EK}) and
(\ref{EK2}) are equal for small field intensities and small wave
vectors.
\section{DISCUSSION AND CONCLUSIONS} Finalizing the paper, we
have to estimate the effects discussed above. The distance between
neighboring electron-photon states (\ref{Energy}) is characterized
by the two energies, $\hbar\omega_0$ and
$\varepsilon_g=\hbar\omega_g$. Therefore the destructive influence
of scattering processes on quasiparticles can be neglected if
$\omega_0\tau\gg1$ and $\omega_g\tau\gg1$, where
$\tau\sim10^{-12}$c is mean free time of charge carriers in
graphene \cite{CastroNeto_09}. Thus the developed concept of
quasiparticles is applicable if the field (\ref{A}) is both
high-frequency and strong. Particularly, such a field can be
created by using lasers. Let the wave is generated by a laser with
the wavelength $\lambda$ and focused into a narrow beam with the
diameter $d\approx\lambda$. Then the wave amplitude is
$E_0\approx4\sqrt{P/c\lambda^2}$, where $P$ is output power of the
laser. Within this scheme the condition of metal-insulator
transition (\ref{MIT}) can be satisfied, for instance, by using an
ordinary low-power CO$_2$-laser with $\lambda=10.6\,\,\mu$m and
$P\approx76$~Watts. In this case the gap (\ref{Gap}) is
$\varepsilon_g\approx117$~meV, that allows to observe the
metal-insulator transition at room temperatures. As to the effect
of THz emission from irradiated graphene, it takes place for
substantially smaller powers $P\sim10^{-2}$~W.

As it follows from the estimations given above, the considered
phenomena can be observable in graphene under a {\it low-power}
irradiation. This is surprising, since a {\it high-power} laser
radiation is usually necessary to observe effects of strong
electron-photon coupling \cite{Scully_b01,Cohen-Tannoudji}. Let us
give simple physical reasons to clarify this unexpected result.
From the quasi-classical viewpoint, the characteristic energy of
electron coupling to circularly polarized electromagnetic wave is
the kinetic energy of electron rotational motion induced by the
wave. In the case of an usual condensed matter with parabolic
electron energy spectrum, $\varepsilon(k)=\hbar^2k^2/2m^\ast$,
this kinetic energy is $W_0^\ast=(eE_0/\omega_0)^2/2m^\ast$. As to
graphene, due to its linear electron energy spectrum,
$\varepsilon(k)=\pm\hbar v_F|k|$, the energy of the electron
rotational motion induced by the wave is $W_0=2v_FeE_0/\omega_0$.
Since we have $W_0/W_0^\ast\gg1$ for weak fields $E_0$, the
electron coupling to a low-power circularly polarized
electromagnetic radiation is substantially stronger in graphene
than in other condensed matters. The observability of the
discussed effects for a low-power laser pumping is crucial from
experimental viewpoint, since high-power lasers fluidize a crystal
lattice and are inapplicable to condensed-matter experiments. Thus
graphene gives an unique opportunity to observe effects of strong
electron-photon coupling, which cannot be observed in other
condensed matters. As a result, the presented theory opens a
significant new area of graphene-related research, where condensed
matter physics and quantum optics meet.

We have demonstrated that the Dirac electron spectrum of graphene
leads to the strong electron interaction with circularly polarized
photons for low light intensities. This results in bound
electron-photon states (charged quasiparticles) which should be
considered as a substantially new kind of field-matter coupling.
The energy spectrum of the quasiparticles is of dielectric type,
that leads to the photon-induced metal-insulator transition. From
applied viewpoint, this effect can be used as a basis for creating
tunable optoelectronic devices.

\begin{acknowledgments}
The work was partially supported by the Seventh European Framework
Programme (Grant FP7-230778), the Russian Foundation for Basic
Research (Grants 10-02-00077 and 10-02-90001), the Russian
Ministry of Education and Science, and ISTC Project B-1708.
\end{acknowledgments}

\end{document}